\documentstyle[emulateapj,epsf]{article}
\begin{document}
\slugcomment{Accepted by The Astrophysical Journal Letters}
\lefthead{Seager, Sasselov \& Scott}
\righthead{A New Calculation of the Recombination Epoch}

\def\heion{\ion{He}{2}}
\hyphenation{an-is-o-tro-pies}
\hyphenation{an-is-o-tro-py}

\title{A New Calculation of the Recombination Epoch}
\author{S. Seager\footnote{Astronomy Department, Harvard University, 
60 Garden Street, Cambridge, MA 02138},
D. D.~Sasselov$^1$
D. Scott\footnote{Department of Physics \& Astronomy, University of
British Columbia, Vancouver, BC, V6T 1Z1}}


\begin{abstract}
We have developed an improved recombination calculation of H, 
\ion{He}{1}, and \ion{He}{2} in the early Universe
which involves a line-by-line treatment of each atomic level. 
We find two major differences compared with previous calculations.
Firstly, the ionization fraction $x_{\rm e}$
is approximately 10\%
smaller for redshifts $\lesssim800$, due to non-equilibrium processes in the excited
states of H. Secondly, \ion{He}{1} recombination
is much slower than previously thought, and is 
delayed until just before H recombines.
We describe the basic physics behind the new results and present a
simple way to reproduce our calculation.  This should enable fast
computation of the ionization history (and quantities such as the 
power spectrum of CMB anisotropies which depend on it)
for arbitrary cosmologies, without the need to consider the hundreds of
atomic levels used in our complete model.
\end{abstract}

\section{Introduction}
We have recently revisited the calculation of the recombination epoch
of the early Universe by making as few approximations as possible and trying
to retain full accuracy (Seager, Sasselov, \& Scott 1999, hereafter Paper~I).
We were motivated by the potential
to measure (with the MAP and Planck satellites) CMB anisotropies at the roughly
1\% level over a wide range of angular scales, and
indications (Hu et al.~1995, hereafter HSSW) that the existing solution for
hydrogen and helium recombination has uncertainties at that level.
Indeed, we uncovered a number of minor improvements,
as well as two more major effects which change the ionization history in
a significant way.

Detailed understanding of the recombination process is crucial for
modeling the power spectrum of CMB anisotropies. Since the seminal
work
of the late 1960s (Peebles 1968, Zel'dovich et al.~1968), several
refinements have been introduced (see discussion in HSSW and Paper I),
but little has changed.
Modern codes for evolving the ionization fraction $x_{\rm e}= n_{\rm
e}/n_{\rm H}$
(where $n_{\rm e}$ is the number density of electrons and $n_{\rm H}$ is the
total number density of H nuclei) have been
based almost entirely on the single differential equation introduced 30
years ago, with a more accurate recombination coefficient, but no
other basic improvement.
 
With today's computing power, there is not the necessity for making
sweeping approximations that existed 30 years ago. We believe our work
represents the most accurate picture to date of how exactly the
Universe as a whole became neutral. In this Letter we summarize 
what is new in the physics involved and present an approximate
treatment of our set of equations which accurately reproduces our
complete calculation. We discuss in detail the physical basis for
each approximation and pay attention to the limits of validity.
The paper is supplemented by a computer code ({\tt recfast}) which can be used
to do the calculation, and in conjunction with, e.g. {\tt cmbfast}
by Zaldarriaga, Spergel \& Seljak (1997), to compute accurate CMB power spectra
for different cosmologies. 

\section{How the Universe Became Neutral}
\label{sec-species}
\subsection{Our Multi-level Calculation}
In the canonical Hot Big Bang picture, the recombination epoch
is when the Universe became cool
enough for protons to capture electrons and form neutral hydrogen. This
recombination process was not instantaneous, because
the electrons, captured into different atomic energy levels, could not
cascade instantaneously down to the ground state. The electrons were
impeded because of fast reionizations out of excited states
due to the huge reservoir of low energy photons, and
because of the high optical depth of the Lyman lines and continuum 
transitions to ground state. Any Lyman line or continuum transition to the
ground state emitted a photon with energy
where there were few blackbody photons, which immediately photoexcited or
photoionized a neighbouring atom in the ground state. Atoms reached the
ground state through either
cosmological redshifting of the Ly$\,\alpha$ line photons, or by the
$2s$--$1s$ two photon process. Because these rates from n=2 to the ground
state were much slower than the net recombination rate to n=2, a
``bottleneck'' occured which slowed down the entire recombination
process.
The Universe
expanded and cooled faster than recombination could be completed, and a
small fraction of free electrons and protons remained. This fraction,
during and after recombination, affects the CMB anisotropies through
the precise shape of the thickness of the photon last scattering
surface (i.e.~the visibility function). 

The `standard' methodology considers an 
`effective 3-level atom' with a ground state, first excited state (n=2)
and continuum, with the states ${\rm n}>2$ represented by a
recombination coefficient. A single ordinary
differential equation can then be derived to describe the ionization
fraction,
equation~(\ref{eq:newstandard_xe}) (See Peebles, 1968, 1993). Many
assumptions go into this derivation,
including: that H excited states are in equilibrium with the
radiation; that stimulated deexcitation is negligible for the
Ly$\,\alpha$ transition;
that a simple recombination coefficient can be used;
that every net recombination results in a ground state atom, so that
the ground state number
density $n_1 = n_{\rm H} - n_{\rm p}$ (where $n_{\rm H}$ is the total
hydrogen number density, ionized or neutral, while $n_{\rm p}$ is the
proton number density, i.e.~ionized H);
that the Ly$\,\alpha$ redshifting can be dealt with
using a simple escape probability; that collisional processes are negligible;
and that He can be ignored.

Our new methodology, made possible with modern computing power, is to
calculate recombination with as few approximations as possible.  
Instead of
evolving a single ODE for $dx_{\rm e}/dt$ for H recombination, we evolve one ODE for each of
300 atomic energy level populations, as well as one for electrons and
one for the matter temperature. 
With this level-by-level treatment, we include thousands of
bound-bound transitions which correctly couple the ODEs, and we
calculate recombination to each atomic level as it evolves with redshift which
eliminates the need for an effective recombination coefficient ($\alpha$ in
\S3). We also
included other effects (which turned out to be negligible), such as
collisional transitions, feedback of distortions to the radiation
field, complete heating and cooling terms, and H chemistry. See Paper
I for full details on this method.
Our new method tests the approximations used in deriving the standard ODE, and
we find all of them to be valid, except for the equilibrium
assumption. The causes and consequences of this are discussed in
\S\ref{sec-Hphysics},
and the effect can be
approximated using the single ODE approach, with a small modification
that artificially speeds up the recombination at low $z$, as discussed
in \S\ref{sec-equations}.

With larger ionization potentials, \ion{He}{1} and \ion{He}{2} recombined
before H.  \ion{He}{2} recombination shows no deviation from previous calculations,
in that we find only small differences compared with Saha equilibrium.  However,
\ion{He}{1} recombination is important for the CMB anisotropies (HSSW), and 
to some extent for the chemistry in the early Universe (Stancil et al.~1999),
and requires closer scrutiny. Our code evolves \ion{He}{1} and \ion{He}{2}
in the same level-by-level method describe above, and simultaneously
with H. Previous calculations for \ion{He}{1} used the Saha
equilibrium method (e.g. HSSW) or an effective 3-level method
(Matsuda, Sato \& Takeda~1971).
We find that the excited states of \ion{He}{1} remain in equilibrium
with the radiation (unlike H), but
that accurate treatment of \ion{He}{1} results in a much delayed
recombination. This can also be reproduced using a simple ODE for
\ion{He}{1}.

\subsection{What is New in the Physics}
\subsubsection{Hydrogen}
\label{sec-Hphysics}
Neutral H production is slightly faster when using our multi-level
calculation, compared to the standard calculation (updated with the
most recent parameters (HSSW)), as shown in Figure~\ref{fig:Hrec}.

For the standard equilibrium case, the net bound-bound
rates are zero, and this is an implicit assumption in deriving
equation~(\ref{eq:newstandard_xe}). We find that at 
$z\lesssim1000$, the net bound-bound
rates become different from zero, because
at low temperatures, the cool
blackbody radiation field means there are few photons for 
photoexcitation of high energy
transitions (e.g.~70--10, 50--4 etc.). In this case
spontaneous deexcitation dominates, causing a 
faster downward cascade to the n=2 state. In
other words, once an electron is captured to, say, n=70, it can cascade
down to the n=2 state faster than in the equilibrium case, because few
photons are around to photoexcite it.
In addition, the faster downward cascade rate is faster than the
photoionization rate from the upper state, and one might view this as
radiative decay stealing
some of the depopulation ``flux'' from photoionization. 
Both the
faster downward cascade and the lower photoionization rate contribute
to the faster net recombination rate.

\centerline{{\vbox{\epsfxsize=8cm\epsfbox{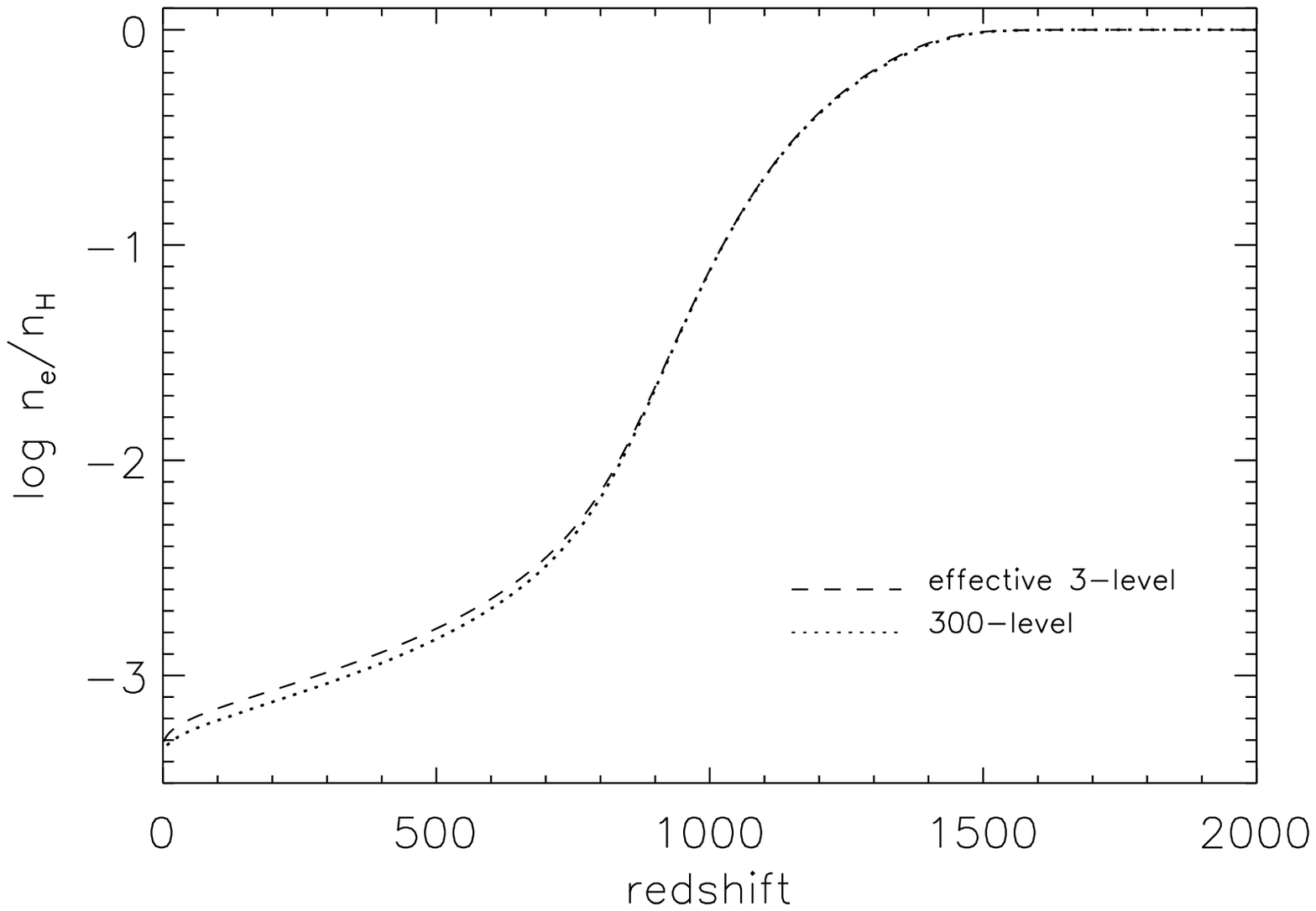}}}}
\noindent{\small
Fig.~1:
Multi-level hydrogen recombination for the standard CDM
parameters $\Omega_{\rm tot} = 1.0, \Omega_{\rm B} = 0.05, H_0 = 50,
Y_{\rm P} = 0.24, T_0 = 2.728\,$K. The two lines are the complete
calculation of Paper~I (dotted) and the standard
(effective 3-level atom) calculation (dashed).
\label{fig:Hrec}
}
\vspace{2mm}

An imbalance develops as the background radiation continues to cool
while the entire Lyman series remains optically thick (the ``bottleneck"
gets worse). A Boltzmann distribution relative to n=2 is no longer
sustainable $-$ the excited states become progressively overpopulated.
(Note that this is not a population inversion.)
The radiation field is cool, but strong, and able to keep neighbouring
states well coupled, as well as to hold the highest Rydberg states in
equilibrium with each other. However, in comparison with the standard
equilibrium capture-cascade calculation for $\alpha$, this unusual
situation leads us to higher effective recombination rates to the majority of
excited states, without increasing photoionization proportionally. This results
in a higher net rate of production of neutral hydrogen atoms.

Because the net effect of our new H calculation is a faster
recombination (a lower freeze out ionization fraction), our results
can be reproduced by artificially speeding up recombination in the
standard calculation, simply by multiplying the recombination and
ionization coefficient by a `fudge factor' $F$. 

\subsubsection{Helium}
Our new multi-level calculation shows a significantly delayed \ion{He}{1}
recombination compared to previous calculations. In fact, for our
low $\Omega_{\rm B}$ models \ion{He}{1} recombination is still
finishing at the very beginning of H recombination. 
Figure~\ref{fig:Herec} shows the ionization fraction $x_{\rm e}$
through \ion{He}{2}, \ion{He}{1} and H recombination, plotted against the
standard H calculation and \ion{He}{2} and \ion{He}{1}
recombination calculated using the Saha ionization equilibrium equation.  
 
\centerline{{\vbox{\epsfxsize=8cm\epsfbox{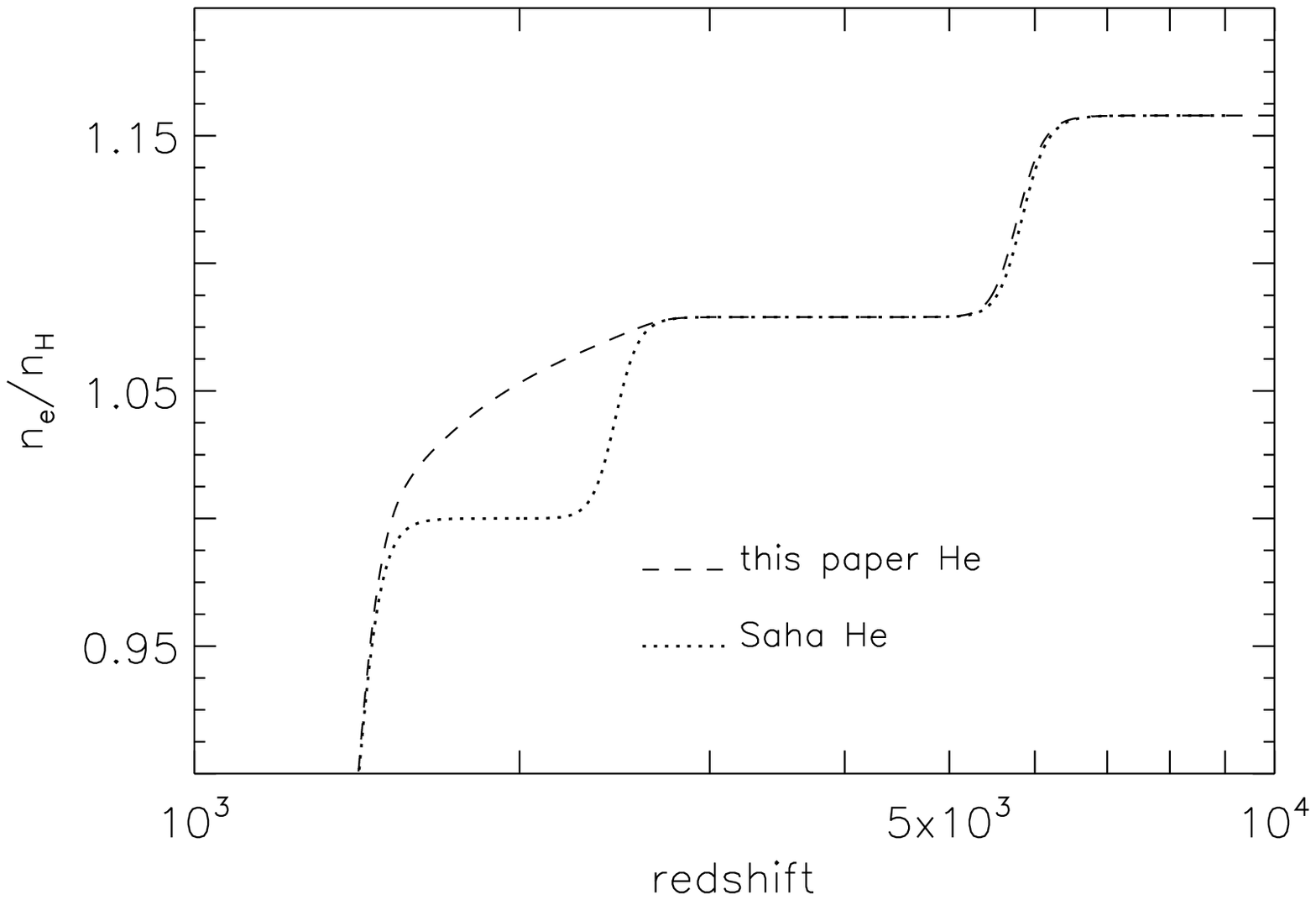}}}}
\noindent{\small
Fig.~2:
Helium and hydrogen recombination for the standard CDM
parameters with $Y_{\rm P} = 0.24$ and $T_0 = 2.728\,$K.  
The first step from right to left is recombination of \ion{He}{3} to
\ion{He}{2}, the second step is \ion{He}{2} to \ion{He}{1}.
The dashed line is our new He I calculation, while the dotted
line is the He Saha equilibrium recombination and the H standard calculation.
\label{fig:Herec}
}
\vspace{2mm}

The recombination of \ion{He}{1} is slow for the same reasons that H
recombination is,
namely because of the optically thick n$^1p$--$1^1s$ transitions which
make cascades to the ground state slow, and the
exclusion of recombinations to the ground state. In other words \ion{He}{1}
follows a Case B recombination with an inhibition factor. Because the
``bottleneck'' at n=2 largely controls recombination, it is not surprising
that \ion{He}{1} and H recombination occur at a similar redshift; the
ionization energy of n=2 is similar in both. 

The physics of \ion{He}{1}
recombination is different than that for H recombination because of
its different atomic structure. 
More specifically, the high excited states of \ion{He}{1} are
much more strongly packed towards the continuum compared to those of
H: the energy difference between the $3p$ levels and the continuum is
$1.6\,$eV for \ion{He}{1} versus $1.5\,$eV for H,
compared to $24.6\,$eV versus $13.6\,$eV for the ground state.
Because of this, the radiation field has a stronger effect on 
the excited states of \ion{He}{1} than those of H. This has 3 main
consequences:
1) The strong radiation field during \ion{He}{1} recombination keeps the
ratio of photoionization rate to downward cascade rate higher than the
H case, resulting in a slower recombination than H. 2) The strong
radiation field also causes the triplet states
(e.g.~n$^3p$--n$^1s$) to be virtually unpopulated.
The triplets have only forbidden (i.e. very slow) transitions with the
singlets, and a metastable ``ground state'' 4.77 eV from the
continuum. 3) The radiation field is
strong enough that the excited states remain
in equilibrium with the radiation throughout recombination. 

There is one possibility to speed up
the slow He I recombination. It is the existence of some
neutral H, which could `steal'  \ion{He}{1} resonance line photons,
invalidate the effective Case~B by removing the ``bottleneck'' at n=2,
and make it a Saha equilibrium recombination
instead. However, our detailed calculation
shows that the rate for this process never comes close
to being significant.

Because \ion{He}{1} recombination simply follows a Case B
recombination and unlike for H the excited singlet states remain in
equilibrium
throughout recombination, it can be reproduced using the effective
3-level single ODE (shown below). Because the triplet excited
states remain unpopulated during recombination, the effective 3-level
atom is the ground state $1^1s$, the first singlet excited state, and the
continuum.

\section{The Approximate Set of Equations}
\label{sec-equations}
Here we present a set of equations which allow a simple way to reproduce
our new recombination results from the multi-level code. Though approximate,
the set provided below includes more of the recombination physics than the
`standard' calculation, and reproduces approximately (via parametrizations) the
departures from equilibrium in H, and the slow \ion{He}{1}
recombination found in Paper~I. 

However, a word of caution is necessary -- the set of equations below have a
range of validity which corresponds roughly to the choice of typical
cosmologies
discussed in Paper~I.  Attempts to calculate recombination in more extreme
cosmologies, or with the addition of extra physics, will probably require
evolution of the full set of equations of the multi-level code.
In addition, our approximations here are designed for optimal use with
the CMB anisotropies, and do not fit as well the range below $z \leq 300$,
where molecular formation becomes important. Therefore detailed study of
the chemistry in the early universe will probably also require
evolution of the full set of equations of the multi-level code.

The approximate set of equations below --- two ODEs for the ionization 
fractions of H and \ion{He}{1}, and an equation for the matter
temperature ---
should be solved simultaneously.
We recommend leaving out \ion{He}{2} recombination entirely, since it
has no effect on the power spectrum of CMB anisotropies. Otherwise
Saha equilibrium is an adequate approximation.
The two ODEs derive from consideration of detailed balance in
the effective 3-level atoms of H and \ion{He}{1} (see Peebles, 1968, 1993).
The total recombination coefficients in each case have been parametrized
to reproduce our multi-level results.
The equation of the total rate of change of the matter temperature is a
truncated version of Paper~I eq.~(69), including adiabatic and Compton
cooling terms. The set is as follows:

\begin{eqnarray}
\label{eq:newstandard_xe}
{dx_{\rm p}\over dz} &= \left(x_{\rm e}x_{\rm p} n_{\rm H} \alpha_{\rm H}
 - \beta_{\rm H} (1-x_{\rm p})
   {\rm e}^{-h\nu_{H2s}/kT_{\rm M}}\right) \\
 &\times\quad{\left(1 + K_{\rm H} \Lambda_{\rm H} n_{\rm H}(1-x_{\rm p})\right)
    \over H(z)(1+z)\left(1+K_{\rm H} (\Lambda_{\rm H} + \beta_{\rm H})
     n_{\rm H} (1-x_{\rm p}) \right)},\nonumber
\end{eqnarray}

\begin{equation}
\label{eq:HeI_xe}
{dx_{\rm He II}\over dz} =
   \left(x_{\rm He II}x_{\rm e} n_{\rm H} \alpha_{\rm HeI}
   - \beta_{\rm HeI} (f_{\rm He}-x_{\rm He II})
   {\rm e}^{-h\nu_{HeI2^1s}/kT_{\rm M}}\right)
\end{equation}
\begin{displaymath}
 \times {\left(1 + K_{\rm HeI} \Lambda_{\rm He} n_{\rm H}
  (f_{\rm He}-x_{\rm He II}){\rm e}^{-h\nu_{ps}/kT_{\rm M}})\right)
  \over H(z)(1+z)\left(1+K_{\rm HeI}
  (\Lambda_{\rm He} + \beta_{\rm HeI}) n_{\rm H} (f_{\rm He}-x_{\rm He II})
  {\rm e}^{-h\nu_{ps}/kT_{\rm M}}\right)},
\end{displaymath}
\begin{equation}
\alpha_{\rm H} = F 10^{-19}\frac{at^{b}}{1 + ct^{d}} \,  \mathrm{m^{3}s^{-1}},
\end{equation}

\begin{equation}
\label{eq:hecoefficient}
\alpha_{\rm HeI} =q\left[\sqrt{T_{\rm M}\over T_2}\left(1+\sqrt{T_{\rm
 M}\over T_2}\right)^{1-p}
 \left(1+\sqrt{T_{\rm M}\over T_1}\right)^{1+p}\right]^{-1}\!
   \mathrm{m^{3}s^{-1}},
\end{equation}

\begin{equation}
\label{eq:cooling}
 \frac{dT_{\rm M}}{dz} = \frac{8\sigma_{\rm T}a_{\rm R}
   T_{\rm R}^4}{3H(z)(1+z)m_{\rm e}c}\,
  \frac{x_{\rm e}}{1+f_{\rm He}+x_{\rm e}}\,(T_{\rm M} - T_{\rm R})
  + \frac{2T_{\rm M}}{(1+z)}. \nonumber
\end{equation}

Equation~(\ref{eq:cooling}) is for the matter temperature, which we recommend
to use in the entire calculation above, due to the small but important effect
resulting from the difference between $T_{\rm M}$ and $T_{\rm R}$ at
low redshift.
Below we list all constants and parameters which appear in the above
equations.

The constants are Boltzmann's constant $k$, Planck's constant
$h$, the speed of light $c$, Thomson scattering cross section
$\sigma_{\rm T}$, the electron mass $m_{\rm e}$, and the radiation constant
$a_{\rm R}$.
The three independent variables are proton fraction $x_{\rm p} =
n_{\rm p}/n_{\rm H}$, singly ionized helium fraction $x_{\rm HeII} =
n_{\rm HeII}/n_{\rm H}$, and matter temperature $T_{\rm M}$.
The dependent variable is
the electron fraction $x_{\rm e} = n_{\rm e}/n_{\rm H} =
x_{\rm p} + x_{\rm HeII}$. Here $n$ refers to number
density, and $n_{\rm H}$ is the total hydrogen number density.

Turning to the atomic data, the H Ly$\,\alpha$ rest wavelength is $\lambda_{{\rm H}2p} =
121.5682\,$nm. The H $2s$--$1s$ frequency
$\nu_{{\rm H}2s} = c/\lambda_{{\rm H}2p}$, is
close enough to Ly$\,\alpha$ that the same averaged wavelength value can
be used. The \ion{He}{1} $2^1p$--$1^1s$ 
wavelength is $\lambda_{{\rm HeI}2^1p} = 58.4334\,$nm. Note that the
\ion{He}{1}
$2^1s$--$1^1s$ frequency $\nu_{{\rm He}I2s} = c/60.1404\,$nm.
Unlike for H, the
separation of \ion{He}{1} $2^1p$ and $2^1s$ is large enough that
$\lambda_{{\rm He}2^1p}$ and $\lambda_{{\rm He}2^1s}$
must be distinguished, hence the extra exponential term in
equation~(\ref{eq:HeI_xe}) over equation~(\ref{eq:newstandard_xe}), with
$\nu_{{\rm He}I2^1p2^1s} =\nu_{{\rm He}I2^1p} - \nu_{{\rm
He}I2^1s}\equiv\nu_{ps}$ (where we define this last notation to save space).
The H $2s$--$1s$ two photon rate is
$\Lambda_{\rm H} = 8.22458\,{\rm s}^{-1}$ (Goldman, 1989), while the
\ion{He}{1}
$2^1s$--$1^1s$ two photon rate is $\Lambda_{\rm He} = 51.3\,{\rm s}^{-1}$
(Drake et al. 1969).

Further, $\alpha_{\rm H}$ is the Case B recombination
coefficient for H, from Hummer (1994), and fitted by
P{\'e}quignot et al.~(1991),
with $a=4.309$, $b=-0.6166$, $c=0.6703$, $d=0.5300$ and
$t= T_{\rm M}/10^{4}\,$K. The factor $F$ is 1.14,
and allows equation~(\ref{eq:newstandard_xe}) to agree with
our multi-level calculation by speeding up recombination.
Note that it also enters into the coefficient $\beta$ via $\alpha$ as
described below. 
$\alpha_{\rm He}$ is the Case B \ion{He}{1} recombination coefficient
for singlets,
from Hummer and Storey (1998). 
The parameters are
$q=10^{-16.744}$, $p=0.711$, $T_1=10^{5.114}\,$K, and $T_2$ fixed
arbitrarily at $3\,$K. This fit is good to $<0.1\%$
over the relevant temperature range ($4{,}000$--$10{,}000\,$K), and still
fairly accurate over a much wider range of temperatures.
The $\beta$ are photoionization coefficients and
are calculated from the recombination coefficients by:
$\beta=\alpha (2\pi m_{\rm e} k \
T_{\rm M}/h^2)^{3/2}
\exp(-h\nu_{2s}/kT_{\rm M})$. Here
$\nu_{2s}$ and $\alpha$ are different for H and \ion{He}{1}. 
Note that $T_{\rm M}$ and
$\nu_{2s}$ are used here, and that incorrectly using $T_{\rm R}$ or
$\nu_{2p}$ will cause a small but important difference for high baryon
models. 

The cosmological parameters are the redshift $z$, Hubble factor
$H(z)$, the cosmological redshifting of H
Ly$\alpha$ photons $K_{\rm H}\equiv\lambda_{\rm H_{2p}}^3/(8\pi H(z))$, the
cosmological redshifting of \ion{He}{1} $2^1p$--$1^1s$ photons
$K_{\rm HeI} \equiv\lambda_{\rm HeI_{2^1p}}^3/(8\pi H(z))$,
 and the radiation temperature
$T_{\rm R} = T_0 (1+z)$.
The primordial He abundance was taken to be $Y_{\rm P}=0.24$ by mass
(Schramm \& Turner, 1998),
and the present-day CMB temperature $T_{0}$ to
be $2.728\,$K (the central value determined by the FIRAS experiment,
Fixsen et al.~1996).

A word of caution about the numerical computation. The equations are
generally stiff, and so there are two approaches
which can be taken. One is to use an integrator appropriate for stiff
sets of equations, the other is to use Saha
equilibrium values of $x_{\rm e}$ to carry the
integrations through the stiffest regimes (the beginning of each recombination
epoch). The former case is not much slower, and certainly will work
for arbitrary cosmology.
For the latter case some experimentation may be necessary in order to
efficiently choose
the redshift to start and finish the Saha approximation for particular
cosmologies, and then it may be faster.
 
For \ion{He}{2} it is sufficient to use the relevant Saha equation:
\begin{equation}
\label{eq:hetwosaha}
{(x_{\rm e}-1-f_{\rm He})x_{\rm e}\over 1+2f_{\rm He}-x_{\rm e}} =
 {(2\pi m_{\rm e} k T)^{3/2}\over h^3 n_{\rm H}}
 {\rm e}^{-\chi_{\rm He II}/kT}.
\end{equation}
For the non-stiff integrator approach, this can be switched on at say
$z>5{,}000$.  Then there will be
a period when He is all singly ionized, and after say $z=3500$ the
Saha equation for \ion{He}{1} needs to be switched on:
\begin{equation}
\label{eq:heonesaha}
{(x_{\rm e}-1)x_{\rm e}\over 1+f_{\rm He}-x_{\rm e}} =
 4 {(2\pi m_{\rm e} k T)^{3/2}\over h^3 n_{\rm H}}
 {\rm e}^{-\chi_{\rm He I}/kT}.
\end{equation}
This can be used until say 1\% of the \ion{He}{1} has recombined, at which
point the coupled ODEs can be solved using a routine such as {\tt
dverk} (Hull et al., 1976).
It is also worth fixing the hydrogen to be fully ionized until some
redshift, then using Saha for H to get recombination started.
It may also save some integration time if He recombination is switched
off once the fraction of neutral He has fallen below some small number.
 
We found that in fact it was not necessary to evolve the H and He
equations simultaneously, at least for cosmologies which we checked in
detail. It appears to be sufficient to evolve each separately, even in cases
where there is a small overlap in the recombination epochs.  However, there
is little computational expense in solving the 3 ODEs simultaneously, and
so that is what we recommend.

\section{Conclusion}
We have presented the basic physics behind our improved recombination
calculation which shows a significantly delayed \ion{He}{1}
recombination and a 10\% lower residual $x_e$ at freezeout compared to
previous calculations. We present
a set of equations to reproduce our work, which are
modified versions of those previously used because
our new, detailed calculation agrees very well
with the results of the standard calculation. This underscores the
tremendous achievement of Peebles, Zel'dovich and colleagues in so
fully understanding cosmic recombination 30 years ago.  However, the
goal of modern cosmology is to determine the cosmological
parameters to an unprecedented level of precision, and in order to do
so it is now necessary to understand very basic things, like
recombination, much more accurately.

\acknowledgments
We would like to thank George Rybicki, Ian Dell'Antonio, Avi Loeb and
Han Uitenbroek for many useful
conversations.  Also David Hummer and Alex Dalgarno for discussions on the
atomic physics, and Jim Peebles for discussions on several aspects of
this work. We thank Martin White, Wayne Hu and Uro{\v s} Seljak for
insisting on this Letter, and the referee for helpful comments.
DS is supported by the Canadian Natural Sciences and Engineering
Research Council.

\end{document}